\documentclass[reprint,amsmath,amssymb,aps]{revtex4-1}

\usepackage{bm}

\begin{document}

\preprint{APS/123-QED}

\title{Acceleration without Temperature}

\author{Alaric D. Doria}
\email{alaricdoria@mail.fresnostate.edu}
\author{Gerardo A. Mu\~noz}%
 \email{gerardom@csufresno.edu}
 \affiliation{Physics Department, California State University, Fresno.}
\date{\today}

\begin{abstract}
We show that while some non-uniformly accelerating observers (NUAOs) do indeed see a Bose-Einstein distribution of particles for the expectation value of the number operator in the Minkowski vacuum state, the density matrix is non-thermal and therefore a definition of temperature is not warranted. This is due to the fact that our NUAOs do not see event horizons in the spacetime. More specifically, the Minkowski vacuum state is perceived by our NUAOs as a single-mode squeezed state as opposed to the two-mode squeezed state characteristic of uniformly accelerating observers. Both single and two-mode squeezed states are pure quantum states; however, tracing over degrees of freedom in one of the modes of the two-mode squeezed state reduces the pure density matrix to a thermal density matrix. It is this property in the two-mode squeezed state that allows one to consistently define a temperature. In the single-mode case, an equivalent tracing is neither required nor available.
\end{abstract}
\maketitle

\section{Introduction}

Soon after the discovery by Hawking \cite{Hawking} that black holes possess temperature, Davies \cite{Davies} and Unruh \cite{Unruh}, following the work of Fulling \cite{Fulling}, showed that observers moving with constant proper acceleration in flat spacetime see the Minkowski vacuum as a thermal bath with temperature $T = \hbar a / 2\pi k $ \cite{Davies,Unruh}. The Unruh effect has deservedly become the prime flat-spacetime example illustrating a deep connection between relativity, quantum mechanics, and thermodynamics. 

In this paper we show that Rindler observers (i.e., observers moving in flat spacetime with constant proper acceleration) are rather special in their identification of the Minkowski vacuum as a thermal state. We make use of the non-uniformly accelerating observer (NUAO) solution found in \cite{DM}, where observers accelerate to terminal velocities less than the speed of light, to study the properties of the Minkowski vacuum seen by this accelerating observer. 

In Sec. II we give a brief review of quantum field theory in non-inertial reference frames, followed by a detailed calculation of the creation and annihilation operators seen by our NUAOs. We show that these accelerating observers see a Bose-Einstein distribution of particles, which strongly suggests the definition of a pseudo-temperature in analogy with the Unruh temperature. 

In Sec. III, we justify that this is indeed a {\sl pseudo}-temperature. We solve for the complete expansion of the Minkowski vacuum in terms of the accelerating observer's particle number states. Furthermore, we prove that our NUAOs experience the Minkowski vacuum as a single-mode squeezed state and not the typical two-mode squeezed state seen by Rindler observers. This single-mode pure quantum state does not allow for a natural reduction to a mixed state, and the thermal state characteristic of the Unruh effect never enters the physics detected by the NUAO. The absence of a thermal state invalidates the seemingly natural definition of temperature suggested by the Bose-Einstein distribution of particles measured by the NUAO.

Sec. IV is concerned with the differences between that of a Rindler observer and our NUAO. We discuss the physics of our NUAOs in the context of squeezed states, entanglement, black hole physics, and the physical implications thereof. A final remark on general accelerations is given.

\section{Quantum Fields and Non-uniformly Accelerating Observers}

We begin with a massless scalar field and show that non-uniformly accelerating observers see a Bose-Einstein distribution for the expectation value of the number operator $\langle  N_\Omega\rangle $ in the Minkowski vacuum. 
\begin{equation}
\label{waveM}
\frac{\partial^2 \hat{\Phi}}{\partial t^2} - \frac{\partial^2 \hat{\Phi}}{\partial x^2} = 0
\end{equation}
We change to light cone coordinates $u = t-x$ and $v = t+x$. The wave equation becomes
\begin{equation}
\label{waveuv}
\frac{\partial^2 \hat{\Phi}}{\partial u \partial v} = 0
\end{equation}
The general solution to (\ref{waveuv}) is
\begin{equation}
\label{waveuvsol}
 \hat{\Phi} = \hat{\Phi}_1 (u) + \hat{\Phi}_2 (v) 
 \end{equation}
 The left $(\hat{\Phi}_1(v))$ and right $(\hat{\Phi}_2(u))$ moving solutions to the field equation are non-interacting; for brevity, we analyze the left-moving solution $\hat{\Phi}_2 (v) = \hat{\Phi}(v)$. We expand the field in terms of Fourier modes as follows:
\begin{equation}
\label{phiv}
\hat{\Phi}(v) = \int_{0}^{\infty} \frac{d\omega}{\sqrt{4\pi \omega}} \left[ \hat{a}^{\vphantom{\dagger}}_\omega  \, e^{-i\omega v} + \hat{a}^\dagger_\omega \, e^{i\omega v} \right]
\end{equation}
where the creation $(\hat{a}^\dagger_\omega)$ and annihilation $(\hat{a}^{\vphantom{\dagger}}_\omega)$ operators for the Minkowski observer obey the standard (discretized) commutation relation $[\hat{a}^{\vphantom{\dagger}}_\omega, \hat{a}^\dagger_{\omega^\prime}] = \delta_{\omega\omega^\prime}$.

The non-uniformly accelerating observer under consideration sees the following metric \cite{DM}:
\begin{equation}
ds^2 = \cosh\left(\frac{X+T}{\alpha}\right)\cosh\left(\frac{X-T}{\alpha}\right)(dT^2 - dX^2)
\end{equation}
where the parameter $\alpha$ is related to the acceleration and the initial velocity of the observer. If we switch to light cone coordinates $U = T-X$ and $V = T+X$ in the accelerating frame, the metric becomes
\begin{equation}
ds^2 = \cosh\left(\frac{V}{\alpha}\right) \cosh\left(\frac{U}{\alpha}\right) dVdU
\end{equation}
The relationship of $U$, $V$ to the corresponding Minkowski coordinates $u$, $v$ is given by~\cite{DM}
\begin{equation}
\label{UV}
u = \alpha \sinh\left(\frac{U}{\alpha}\right)  \quad  ,  \quad  v = \alpha \sinh\left(\frac{V}{\alpha}\right) 
\end{equation}
Note that both coordinate systems $(u,v)$ and $(U,V)$ cover the entire spacetime and there is no problem with the origin in the accelerating frame \cite{DM}. In other words, the acceleration of the NUAO does not become infinite as we approach the origin. Additionally, this results in a single pair of creation and annihilation operators for the accelerating observer, in contrast to the two pairs required for uniformly accelerating observers. 

The equation satisfied by the scalar field in the accelerating observer's coordinates $U$ and $V$ is formally identical to (\ref{waveuv}).
\begin{equation}
\frac{\partial^2 \hat{\Phi}}{\partial U \partial V} = 0
\end{equation}
The solution is then analogous to (\ref{waveuvsol}). Selecting again the left-moving solution we write
\begin{equation}
\label{phiV}
\hat{\Phi}(V) = \int_{0}^{\infty} \frac{d\Omega}{\sqrt{4\pi \Omega}} \left[ \hat{b}^{\vphantom{\dagger}}_\Omega \, e^{-i\Omega V} + \hat{b}^\dagger_\Omega \, e^{i\Omega V} \right]
\end{equation}
in terms of the Fourier modes. The creation ($\hat{b}^\dagger_\Omega$) and annihilation ($\hat{b}^{\vphantom{\dagger}}_\Omega$) operators in the accelerating frame satisfy the same commutation relation $[ \hat{b}^{\vphantom{\dagger}}_\Omega  \, , \hat{b}^\dagger_{\Omega^\prime} ]= \delta_{\Omega\Omega^\prime}$. We identify equations (\ref{phiv}) and (\ref{phiV}) and solve for  the creation and annihilation operators in the accelerating frame. We multiply by $e^{-i\Omega^{\prime} V}$ and integrate over $V$ to obtain $\hat{b}^\dagger_{\Omega}$.
\begin{equation}
\hat{b}^\dagger_\Omega = \frac{\sqrt{\Omega} }{2\pi}\int_{-\infty}^{\infty} \! dV \int_{0}^{\infty} \! \frac{d\omega}{\sqrt{\omega}}  \Big[ \hat{a}^{\vphantom{\dagger}}_\omega  \, e^{-i\omega v} \, e^{-i\Omega V} + \hat{a}^\dagger_\omega \, e^{i\omega v} \, e^{-i\Omega V} \Big]
\end{equation}
We refer to equations (\ref{UV}) and replace $v =\alpha \sinh(V/\alpha)$ in the exponentials inside the integral. The integral formula for the modified Bessel function of the second kind \cite{Watson}
\begin{equation}
K_{\nu}(x) = \frac{1}{2} e^{-i\nu \pi /2} \int_{-\infty}^{\infty} dt \, e^{-ix \sinh t -\nu t}
\end{equation}
is valid when $x > 0$ and $-1< \Re(\nu) <  1$. Therefore, using the fact that $K_{- \nu} = K_{\nu}$, we have

\begin{equation}
\label{regularize}
\hat{b}^\dagger_\Omega = \frac{\alpha}{\pi} \sqrt{\Omega} \int_{0}^{\infty} \frac{d\omega}{\sqrt{\omega}} \: K_{i\Omega \alpha}(\omega \alpha) \left[ e^{-\pi \Omega \alpha / 2} \hat{a}^{\vphantom{\dagger}}_\omega + e^{\pi \Omega \alpha / 2} \hat{a}^\dagger_\omega \right]
\end{equation}
At this point, the integral diverges, of course, and a regularization procedure must be adopted. The details do not matter for our purposes, since our results will be independent of the specific regularization procedure. Denoting by $\chi(\omega)$ the combination
\begin{equation}
\chi(\omega) \equiv  \frac{\alpha}{\pi} \sqrt{ \frac{\Omega} {\omega} } \: K_{i\Omega\alpha}(\omega \alpha)
\end{equation}
we may write the expression for $\hat{b}^\dagger_\Omega$ as
\begin{equation}
\label{bdag}
\hat{b}^\dagger_\Omega = e^{-\pi \Omega \alpha / 2} \int_{0}^{\infty} d\omega \chi(\omega) \, \hat{a}^{\vphantom{\dagger}}_\omega  + e^{\pi \Omega \alpha / 2} \int_{0}^{\infty} d\omega \chi(\omega) \, \hat{a}^\dagger_\omega
\end{equation}
Similarly, taking the adjoint of equation (\ref{bdag}) gives the following expression for $\hat{b}^{\vphantom{\dagger}}_\Omega$.
\begin{equation}
\label{bbdag}
\hat{b}^{\vphantom{\dagger}}_\Omega = e^{\pi \Omega \alpha / 2} \int_{0}^{\infty} d\omega \chi(\omega) \, \hat{a}^{\vphantom{\dagger}}_\omega  + e^{-\pi \Omega \alpha / 2} \int_{0}^{\infty} d\omega \chi(\omega) \,\hat{a}^\dagger_\omega 
\end{equation}

We may now isolate the operator $\hat{a}^{\phantom{\dagger}}_\omega$ that annihilates the Minkowski vacuum $|0_M\rangle$ by solving the system of equations (\ref{bdag}) and (\ref{bbdag}) for $\hat{a}^{\vphantom{\dagger}}_\omega$.
\begin{equation}
e^{\pi \Omega \alpha } \;  \hat{b}^{\vphantom{\dagger}}_\Omega - \hat{b}^\dagger_\Omega  =  \left(  e^{2\pi \Omega \alpha }  -1 \right)  \int_{0}^{\infty} d\omega \chi(\omega) \, \hat{a}^{\vphantom{\dagger}}_\omega
\end{equation}
The combination $ \hat{b}^{\vphantom{\dagger}}_\Omega - e^{-\pi \Omega \alpha } \; \hat{b}^\dagger_\Omega$ therefore annihilates the Minkowski vacuum,
\begin{equation}
\label{bb0}
\left( \hat{b}^{\vphantom{\dagger}}_\Omega - e^{-\pi \Omega \alpha } \; \hat{b}^\dagger_\Omega \right) |0_M \rangle  = 0
\end{equation}
We construct the number operator $N_\Omega = \hat{b}^\dagger_\Omega\hat{b}^{\vphantom{\dagger}}_\Omega$ for the accelerating observer and determine the average particle number $\langle 0_M | \hat{b}^\dagger_\Omega \, \hat{b}^{\vphantom{\dagger}}_\Omega  |0_M \rangle$. Using equation (\ref{bb0}) we find that the norm of the state $\hat{b}^{\vphantom{\dagger}}_\Omega  |0_M \rangle$ is
\begin{equation}
\label{norm}
\langle 0_M | \hat{b}^\dagger_\Omega \, \hat{b}^{\vphantom{\dagger}}_\Omega  |0_M \rangle = \langle 0_M | \, e^{-2\pi \Omega \alpha } \, \hat{b}^{\vphantom{\dagger}}_\Omega  \; \hat{b}^\dagger_\Omega   |0_M \rangle
\end{equation}
Finally, we use the commutator $[ \hat{b}^{\vphantom{\dagger}}_\Omega  \, , \hat{b}^\dagger_\Omega ]= 1$ on the right side of equation (\ref{norm}) to obtain
\begin{equation}
\label{<N>}
\langle 0_M | \hat{b}^\dagger_\Omega \, \hat{b}^{\vphantom{\dagger}}_\Omega  |0_M \rangle = \left( e^{2\pi \Omega \alpha } -1 \right)^{-1}
\end{equation}
Equation (\ref{<N>}) shows that the non-uniformly accelerating observer sees a Bose-Einstein distribution in the Minkowski vacuum. One would therefore expect to be able to define a temperature from $2\pi \Omega \alpha = E/k T$ and $E = \hbar \Omega$ as
\begin{equation}
\label{temp}
T  = \frac{\hbar}{   2\pi k \alpha}
\end{equation}
directly from the Bose-Einstein statistics. Although the association of the state with a temperature seems well justified, we now show that this is not the case.
\section{Expansion of the Minkowski Vacuum}
We expand the Minkowski vacuum state in terms of the accelerating observer's number eigenstates. 
\begin{equation}
\label{vacuumexp}
|0_M\rangle  = \sum_{\lbrace n_i \rbrace}|\lbrace n_i \rbrace\rangle \langle \lbrace n_i \rbrace |0_M \rangle 
\end{equation}
The states $|\lbrace n_i \rbrace\rangle $ are calculated by repeated action of the $\hat{b}^\dagger_\Omega$ creation operators on the accelerating observer's vacuum state $ |0 \rangle $. We abbreviate $ \hat{b}^{\vphantom{\dagger}}_{\Omega_i}  =  \hat{b}^{\vphantom{\dagger}}_i$ and write  $\langle \lbrace n_i \rbrace |0_M \rangle $ in terms of products of annihilation operators acting to the left. 
\begin{equation}
\label{nMvac}
\langle \lbrace n_i \rbrace|0_M\rangle  = \frac{1}{\sqrt{n_1!\cdot n_2! ...}} \langle 0| (\hat{b}_1)^{n_1} (\hat{b}_2)^{n_2}...|0_M\rangle 
\end{equation}
Equation (\ref{bb0}) in combination with repeated applications of the commutator $[ \hat{b}^{\vphantom{\dagger}}_\Omega  \, , \hat{b}^\dagger_\Omega ]= 1 $ yields
\begin{equation}
b^n |0_M\rangle  = e^{-\pi\Omega\alpha} (N + n - 1)\hat{b}^{n-2}|0_M\rangle 
\end{equation}
where $N = b^\dagger b$ is the number operator. Iterating the process for $\hat{b}^{n-2}$, $\hat{b}^{n-4}, \ldots$ we find
\begin{equation}
\label{nMvac2}
\langle \lbrace n_i \rbrace |0_M\rangle   = 
\begin{cases}
\langle 0|0_M\rangle  \prod_{j} e^{-n_j \pi \Omega_j \alpha / 2} \,  \frac{(n_j - 1)!! }{ \sqrt{n_j!} } \quad (n_j \; $even$) \\
\quad 0 \quad \quad \quad \quad \quad \quad \quad \quad \quad \quad \quad \quad \; (n_j \; $odd$)
\end{cases}
\end{equation}
We relabel $n_i$ by $2k_i$ where $k_i$ is the number of pairs of particles; this allows us to write a summation over all $k_i$. The expansion of the Minkowski vacuum becomes
\begin{equation}
\label{Mvacexp}
|0_M\rangle  = \langle 0|0_M\rangle  \sum_{\{k_i\}} \Big(\prod_i \frac{(2k_i-1)!!}{\sqrt{(2k_i )!}}e^{-k_i \pi\Omega_i\alpha}\Big) |\lbrace 2k_i \rbrace\rangle
\end{equation}

We calculate the transition amplitude $\langle 0|0_M \rangle$ by left-multiplying by $\langle 0_M |$, replacing $\langle 0_M |\{2k_i\}\rangle$ with our result from equation (\ref{nMvac2}) and switching the order of product and summation.
\begin{equation}
\label{Mproduct}
1 = \langle 0_M|0_M\rangle  = \vert\langle 0|0_M\rangle \vert^2 \prod_{i} \sum_{k_i=0}^{\infty} e^{- 2k_i \pi \Omega_i \alpha } \; \frac{[(2k_i - 1)!!]^2}{ (2k_i )!}
\end{equation}
Each sum in equation (\ref{Mproduct}) is given by the closed form
\begin{equation}
\label{exactsum}
\sum_{k=0}^{\infty} (e^{-2 \pi \Omega_i \alpha})^k \frac{[(2k - 1)!!]^2}{ (2k)!} = \left(1-e^{-2\pi \Omega_i \alpha}\right)^{-1/2}
\end{equation}
Substitution of the closed form (\ref{exactsum}) into equation (\ref{Mproduct}) leads to
\begin{equation}
\label{Zdef}
\langle 0|0_M\rangle = \prod_{i} \left(1-e^{-2\pi \Omega_i \alpha}\right)^{1/4} \equiv \mathcal{Z}^{-1/2}
\end{equation}
where we have dropped an irrelevant phase in extracting the square root. Replacing this result for the vacuum-to-vacuum transition amplitude into equation (\ref{Mvacexp}) determines the expansion of the Minkowski vacuum in terms of pairs of bosons to be
\begin{equation}
\label{completeexpansion}
|0_M\rangle  = \mathcal{Z}^{-1/2}  \sum_{\{k_i\}} \Big(\prod_i \frac{(2k_i-1)!!}{\sqrt{(2k_i )!}} \, e^{-k_i \pi\Omega_i\alpha}\Big) |\lbrace 2k_i \rbrace\rangle
\end{equation}
We may now elucidate the relationship between the two vacua. Writing equation (\ref{completeexpansion}) as an operator relationship between the two vacuum states is easily accomplished,
\begin{equation}
\label{squeezingoperator}
|0_M\rangle  = \mathcal{Z}^{-1/2} \prod_i \exp \left[\frac{1}{2} \, e^{-\pi\Omega_i\alpha} \left(b_{\Omega_i}^{\dagger}\right)^2\right] |0\rangle
\end{equation}
With some effort, one can show that equation (\ref{squeezingoperator}) is equivalent to
\begin{equation}
|0_M\rangle  = \prod_i \exp \Big\{ -\frac{1}{4} \ln \tanh\left(\frac{\pi \Omega_i \alpha}{2}\right)  \Big[\left(b_{\Omega_i}^{\dagger}\right)^2 - \left(b_{\Omega_i}^{\phantom{\dagger}}\right)^2\Big] \Big\} \, |0\rangle
\end{equation}
Note that the operator acting on $|0\rangle$ is unitary and is in fact the product of squeezing operators for single-mode states \cite{Agarwal}. This shows explicitly that the Minkowski vacuum $|0_M\rangle$ is a single-mode squeezed $|0\rangle$ vacuum, and equation (\ref{<N>}) may now be seen to be in agreement with a well-known result in quantum optics. 

\section{Conclusions}
We have shown that the Minkowski vacuum is perceived as a pure single-mode squeezed state by the accelerating observer. 

In the case of uniformly accelerating Rindler observers, the same Minkowski vacuum is perceived as a pure two-mode squeezed state. However, Rindler observers see an event horizon which forces a tracing, in either the left or right wedge, over inaccessible entangled degrees of freedom. The loss of entangled information reduces the pure density matrix $\rho = |0_M \rangle \langle 0_M|$ to a thermal mixture.

 In the case of the non-uniformly accelerating observer, there is no need for an equivalent tracing over inaccessible degrees of freedom, or even a motivation for doing such a thing given that this observer sees no event horizon and has complete access to the entire spacetime. Therefore, defining a temperature $T = \hbar/2\pi k \alpha$ is not justified despite the fact that the Bose-Einstein nature of the particle number distribution appears to suggest a thermal state. 
 
 Stated differently, thermal states should not be associated with accelerations in general, but only with accelerations capable of introducing horizons in the frame of the accelerating observer. 

 Even constant acceleration does not guarantee the existence of a temperature \cite{DL}.

From the point of view of black hole physics, the observers considered in this paper are in a sense ``intermediate" between freely falling and stationary observers. It would be interesting to see if these observers can shed any additional light on the issues of black hole complementarity and firewalls \cite{AMPS,NVW}.

\section*{Acknowledgement}

We thank S. Deser for bringing Ref. \cite{DL} to our attention.

\section*{Appendix A: Rindler Observers and Unruh Temperature}

We have already noted that equation (\ref{temp}) cannot be interpreted as the temperature of the particle distribution detected by our accelerating observer. Of course, a similar definition {\sl is} warranted in the case of a Rindler observer. This naturally raises the question as to whether the Unruh temperature can be obtained from our results. 

At first sight it may appear that we have two incompatible expressions, since the Rindler observer is the $\alpha \rightarrow 0$ limit of our NUAO. However, as pointed out in \cite{DM}, the limit should be taken as the NUAO's incoming velocity  $ v_{\infty} \rightarrow 1$ with the product $\alpha \gamma_{\infty}$ fixed, where $ \gamma_{\infty}$ is the gamma factor associated with $v_{\infty}$. Furthermore, in this limit the relationship between the observer's time $dT$ and the proper time $d \tau$ becomes $ \gamma_{\infty} dT = d \tau$. 

This implies that energies and temperatures defined with respect to the proper time must include a gamma factor; in particular,
\begin{equation}
\label{temp2}
T_{\tau}  = \frac{\hbar}{  2\pi k \alpha  \gamma_{\infty} }
\end{equation}
replaces equation (\ref{temp}). Since $\alpha \gamma_{\infty} = 1/a_{pr}$, as $v_\infty \rightarrow 1$ \cite{DM}, the temperature becomes
\begin{equation}
\label{temp2}
T_{\tau}  = \frac{\hbar \,  a_{pr}}{  2\pi k}
\end{equation}
and we have agreement with the Unruh temperature.

\section*{Appendix B: Verification of Bose-Einstein Distribution of Particle Numbers}

We check our expansion of the vacuum state by an independent calculation of the expectation value for the number operator. From equation (\ref{completeexpansion}), $\langle 0_M|N_{\Omega_j}|0_M\rangle$ is
\begin{equation}
\langle 0_M| \mathcal{Z}^{-1/2} \sum_{\lbrace k_i \rbrace}\Big(\prod_i \frac{(2k_i-1)!!}{\sqrt{(2k_i )!}} \, e^{-k_i \pi\Omega_i\alpha}\Big)N_{\Omega_j} |\lbrace 2k_i \rbrace\rangle 
\end{equation}
The number operator acting on the states gives $N_{\Omega_j}|\{2k_i\}\rangle = 2k_j |\{2k_i\}\rangle$. Substitution of $\langle 0_M |\lbrace 2k_i \rbrace\rangle$ from (\ref{nMvac2}) then yields
\begin{equation}
\langle 0_M|N_{\Omega_j}|0_M\rangle  =  \mathcal{Z}^{-1} \sum_{\lbrace k_i \rbrace}\Big(\prod_i \frac{(2k_i-1)!!}{\sqrt{(2k_i )!}} \, e^{-k_i \pi\Omega_i\alpha}\Big)^2 2 k_j
\end{equation}
Equations (\ref{exactsum}) and (\ref{Zdef}) imply that the infinite product will produce a cancellation of every $\mathcal{Z}_i$ factor in the product, with a corresponding term in the normalization, except when $i=j$. Hence
\begin{equation}
\langle 0_M|N_{\Omega_j}|0_M\rangle  =  \mathcal{Z}_j^{-1} \sum_{k_j =0}^\infty  \frac{[(2k_j-1)!! ]^2}{(2k_j)!} \, e^{-2k_j \pi\Omega_i\alpha} \; 2 k_j
\end{equation}
The remaining sum is easily evaluated by taking a derivative of (\ref{exactsum}). The result is
\begin{equation}
\langle 0_M|N_{\Omega_j}|0_M\rangle  = (e^{2\pi\alpha\Omega_j}-1)^{-1}
\end{equation}
This is the same Bose-Einstein distribution found in Section 2.

\end{document}